\documentclass[twocolumn,aps,prl,superscriptaddress,tightenlines]{revtex4}
\usepackage{amsmath}
\usepackage{amsfonts}
\usepackage{graphicx}
\usepackage{epsfig}
\usepackage{color}
\usepackage{textcomp}
\usepackage[colorlinks,citecolor=blue]{hyperref}

\begin{document}

\title{Parity-Symmetry-Protected Multiphoton Bundle Emission}
\author{Qian Bin}
\affiliation{School of Physics, Huazhong University of Science and Technology, Wuhan, 430074, P. R. China}

\author{Ying Wu}
\affiliation{School of Physics, Huazhong University of Science and Technology, Wuhan, 430074, P. R. China}

\author{Xin-You L\"{u}}\email{xinyoulu@hust.edu.cn}
\affiliation{School of Physics, Huazhong University of Science and Technology, Wuhan, 430074, P. R. China}

\date{\today}
             
\begin{abstract}
We demonstrate symmetry protected multiphoton bundle emission in the cavity QED system under the ultrastrong coupling regime. Our proposal only enables the super-Rabi oscillations with periodic generation of even correlated photons in the cavity, which is realized by combining the laser driven flip of qubit and the symmetry conserved transitions induced by Rabi interaction with parity symmetry. Combined with dissipation, only 2$n$-photon bundle emissions are allowed, due to the almost perfect suppression of bundle emissions with odd correlated photons. Meanwhile, the corresponding purities are significantly enhanced by the parity symmetry. This work extends multiphoton bundle emission to the ultrastrong coupling regime, and offers the prospect of exploring symmetry-protected multiphoton physics. 
\end{abstract}
\maketitle

Symmetry plays a key role in understanding and exploring the fundamental properties of physical systems\,\cite{H. Weyl}. Typically, the presence of a certain symmetry simplifies the calculation of dynamics by organizing the kinematic space of theory with the irreducible representation of symmetry groups\,\cite{S. Singh}. The cavity quantum electrodynamics (QED) system described by the quantum Rabi model possesses a parity (or $\mathbb{Z}_2$) symmetry. This decides that the system is integrable~\cite{D. Braak2011} and the dynamical evolution is confined to the parity conserved subspace\,\cite{J. Casanova2010}. Under the rotating wave approximation (RWA), the Rabi model is reduced to the Janes-Cummings model\,\cite{Jaynes1963}, where the discrete $\mathbb{Z}_2$ symmetry is extended to a continuous $U(1)$ symmetry. However the RWA is not valid anymore in the ultrastrong coupling regime, where the light-matter coupling rate can reach an order of $10\%$ of the natural frequencies of the noninteracting parts~\cite{G. Gunter,Y. Todorov,P. Forn-Diaz,T. Schwartz,A. J. Hoffman,C. M. Wilson,G. Scalari,S. Gambino,P. Forn-Diaz2,  FornDiaz2019}.  Many novel phenomena and applications have been predicted in this regime\,\cite{Kockum2019, Gu2017}, including the generation of correlated photon pairs\,\cite{C. Ciuti2},  vacuum degeneracy\,\cite{P. Nataf}, the implementation of ultrafast quantum gates\,\cite{Romero2012},  multiphoton Rabi oscillation\,\cite{Garziano2015,Ma2020},  quantum nonlinear optics with virtual photons\,\cite{R. Stassi,Kockum2017}, and parity-sensitive dynamics with two-photon relaxation\,\cite{Malekakhlagh2019}.

Multiquanta physics is an increasingly popular research field in modern quantum science, with important applications in quantum metrology\,\cite{V. Giovannetti, M. DAngelo} and quantum biology\,\cite{W. Denk,N. Horton}. Recently, various multiquanta processes have been studied enormously in atom~\cite{M. F. Maghrebi, K. Jachymski,Q.-Y. Liang,Deng2020}, waveguide~\cite{J. S. Douglas,A. Gonzalez-Tudela2} and cavity~\cite{Y. Ota, C. S. Munoz1,Y. Chang, C. S. Munoz2, W. Qin,  K. J. Satzinger} QED systems. In particular, $n$-photon bundle emission is proposed based on Purcell enhancing the so-called leapfrog transitions\,\cite{C. S. Munoz1}, and it is significant for both fundamental and applied quantum optics~\cite{D. V. Strekalov}. Multiphoton bundle emission means that the quantum emitter releases energy  in bundle of $n$ photons, i.e., the unit of emission is effectively replaced by a the bundle of strongly correlated $n$ photons. The antibunched and uncorrelated bundles can be used to realize the multiphoton source~\cite{Y. W. Chu} and laser~\cite{D. J. Gauthier}, respectively. This is essentially different from the collective radiance~\cite{Q. Bin2} that releases energy in the form of single-photon emission and investigates the enhanced radiance behavior of more than one qubit. However, such multiphoton processes are inevitably disturbed by the off-resonance processes in the strong dissipation regime.  A natural question is whether the symmetry of system could solve this problem, and influence the multiphoton bundle emission significantly. Moreover, the bundle emission theory in the ultrastrong coupling regime is largely unexplored, which may substantially advance the field of multiphoton physics.   

Here, we propose a parity-symmetry-protected multiphoton bundle emission by combining the laser driven qubit flip and the Rabi interaction induced sideband transitions. Physically, the parity symmetry featured by Rabi interaction divides the Hilbert space of dynamical evolution into two symmetry-protected subspaces, i.e., odd and even parity chains. Then, after the laser induced qubit flip $|0,g\rangle\rightarrow|0,e\rangle$, Rabi interaction only allows the states in the odd parity chain to be occupied. This ultimately leads to parity-symmetry-protected super-Rabi oscillation\,\cite{D. V. Strekalov} $|0,g\rangle\leftrightarrow|2n,e\rangle$ when we only consider the laser driven resonant processes under the condition of the qubit frequency being much higher than the cavity frequency, i.e., safely ignoring the direct transitions from $|0, g\rangle$ induced only by the Rabi interaction. Combined with the dissipation of system, the super-Rabi oscillations $|0,g\rangle\leftrightarrow|2n,e\rangle$ transfer coherently pure $2n$-photon states outside of the cavity, while the emissions with odd photons are suppressed perfectly. Consequently, parity symmetry significantly enhances the purity of $2n$-photon bundle emission, and almost ideal two-photon emission is possible with current technologies\,\cite{FornDiaz2019,Kockum2019, Gu2017}. Our work builds the connection between the parity symmetry and the multiphoton bundle emission for the first time. It opens up a door for exploring the crossover between the multiphoton physics and symmetry theory, and offers potential applications for the engineering of new types of quantum devices, such as symmetry-protected multiphoton guns and lasers. 

\begin{figure}
\includegraphics[width=8.5cm]{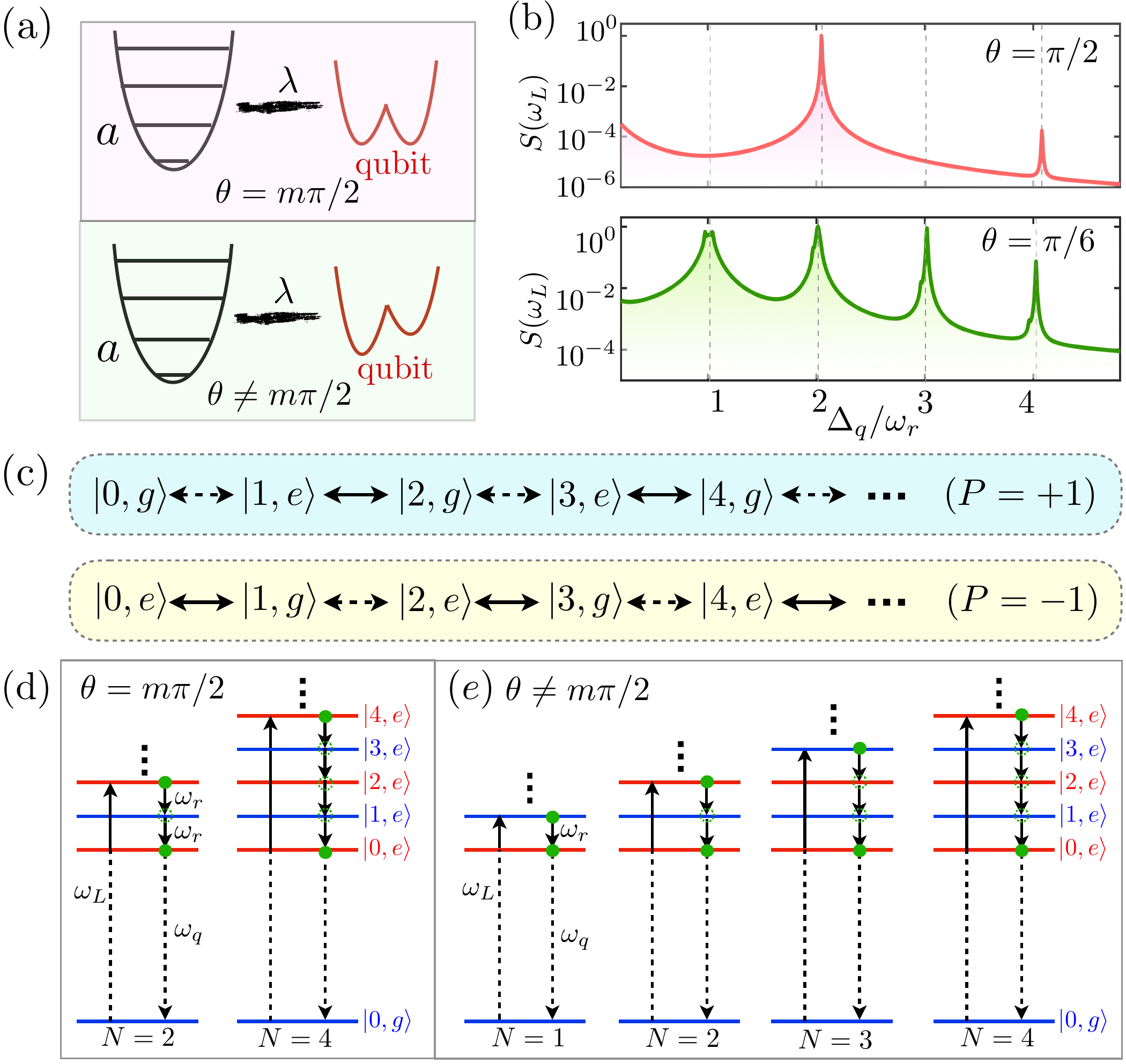}\\
\caption{ (a) Scheme of the model with ($\theta=m\pi/2$) and without ($\theta\neq m\pi/2$) parity symmetry. Here $m$ is the odd number. (b) Photon excitation spectrum for $\theta=\pi/2$ and $\theta=\pi/6$, where $\omega_q/\omega_r=5$,  $\lambda/\omega_r=0.2$, $\Omega/\omega_r=0.06$, $\gamma/\omega_r=0.0001$, and $\kappa/\gamma=20$. (c) Transition between states in the even ($P=+1$) or odd ($P=-1$) parity chain is connected via either rotating (solid arrows) or counterrotating (dashed arrows) terms. (d,e) Multiphoton resonance in two cases. Laser driven qubit flip (black dashed lines) and qubit-cavity interaction induced sideband transitions (black arrows). After a cascade emission of photons and a qubit flip, the system goes back to $|0,g\rangle$. Red and blue lines correspond to the states  with odd and even excitation numbers, respectively.  
 }\label{fig1}
\end{figure}

\emph{Model and multiphoton resonance}.---We consider a cavity QED system, with a qubit ultrastrongly coupled to a single-mode cavity field with coupling strength $\lambda$, as shown in Fig.\,\ref{fig1}(a).  The qubit is driven by a laser with frequency $\omega_L$ and amplitude $\Omega$, and the system Hamiltonian ($\hbar=1$) is $H=\tilde{H}_R+\Omega \cos (\omega_L t) \sigma_x$,
where the extended Rabi Hamiltonian $\tilde{H}_{R}$ reads 
\begin{align}\label{eq2}
\!\tilde{H}_R=\frac{\omega_q}{2} \sigma_z+ \omega_r a^\dag a+ \lambda (\cos\theta \sigma_z-\sin\theta \sigma_x) (a^\dag+a).
\end{align}
Here $a$ ($a^\dag$) is the annihilation (creation) operator of the cavity field, $\sigma_x=\sigma^\dag +\sigma$ ($\sigma=|g\rangle\langle e|$) and $\sigma_z$ are Pauli operators of the qubit. In our proposal, the qubit frequency $\omega_q$ is much higher than the cavity frequency $\omega_r$. The mixing angle $\theta$ decides the relative contribution of the transverse and longitudinal couplings between the qubit and cavity. This extended Rabi Hamiltonian $\tilde{H}_{R}$ can be implemented in the circuit QED system\,\cite{Liu2005, T. Niemczyk,F. Deppe,A. Fedorov}, and it is reduced to the standard Rabi Hamiltonian $H_{R}$ when $\theta=m\pi/2$ ($m$ is any odd number). Interestingly, the Rabi Hamiltonian possesses parity symmetry with a well-defined parity operator $\Pi=e^ {i\pi [a^\dag a+(\sigma_z+1)/2]}$, i.e., $[\Pi,H_R]=0$\,\cite{J. Casanova2010, D. Braak2011}. The eigen-equation $\Pi|P\rangle=P|P\rangle$ ($P=\pm1$) defines the odd-even parity states of system, which constitute two unconnected parity chains of the system evolution, i.e., the odd ($P=-1$) and even ($P=+1$) parity chains shown in Fig.\,\ref{fig1}(c). 

Combining  the laser driven qubit flip and qubit-cavity interaction induced multiphoton sideband transitions, the laser driven resonant processes $|0,g\rangle\leftrightarrow|j,e\rangle$ ($j$ is an integer) are realized, as shown in Figs.\,\ref{fig1}(d) and \ref{fig1}(e). Under the condition of resonantly driving, i.e., $\Delta_q=\omega_L-\omega_q\approx j\omega_r$, the system is first excited to the state $|0,e\rangle$ from $|0,g\rangle$ via qubit flip. For the case of $\theta=m\pi/2$, the Rabi interaction only allows the transition to $|2n,e\rangle$ (i.e., $j=2n$) from $|0,e\rangle$ due to its parity symmetry. This ultimately leads to parity-symmetry-protected super-Rabi oscillation $|0,g\rangle\leftrightarrow|2n,e\rangle$ when we only consider the above dominated resonant processes. The super-Rabi oscillation $|0,g\rangle\to|2n-1,e\rangle$ cannot be allowed in this case, since the transition from  $|0,e\rangle$ to $|2n-1,e\rangle$ is forbidden by the parity symmetry of Rabi interaction after the qubit flip $|0,g\rangle\leftrightarrow|0,e\rangle$. Here we have safely neglected the direct transitions from $|0,g\rangle$ to all occupational states of the cavity induced only by the Rabi interaction, such as $|0,g\rangle\leftrightarrow|1,e\rangle$, which is far off resonant and ignorable in our work\,\cite{supp}. When $\theta\neq m\pi/2$, the parity symmetry of the qubit-cavity interaction is broken, i.e., $[\Pi, \tilde{H}_R]\neq 0$, which enables the connection between two parity chains via the parity-symmetry-breaking transitions, e.g., $\sigma_z a^\dag|n,e\rangle\rightarrow|n+1,e\rangle$. Then, transitions from $|0,e\rangle$ will not be limited in the odd parity chain, and the state $|n,e\rangle$ (i.e., $j=n$) with any integer $n$ can be occupied, as shown in Fig.\,\ref{fig1}(e). The parity-symmetry-protected multiphoton resonances are also shown by the excitation spectrum of Fig.\,\ref{fig1}(b).  

\begin{figure}
  \centering
  \includegraphics[width=8.7cm]{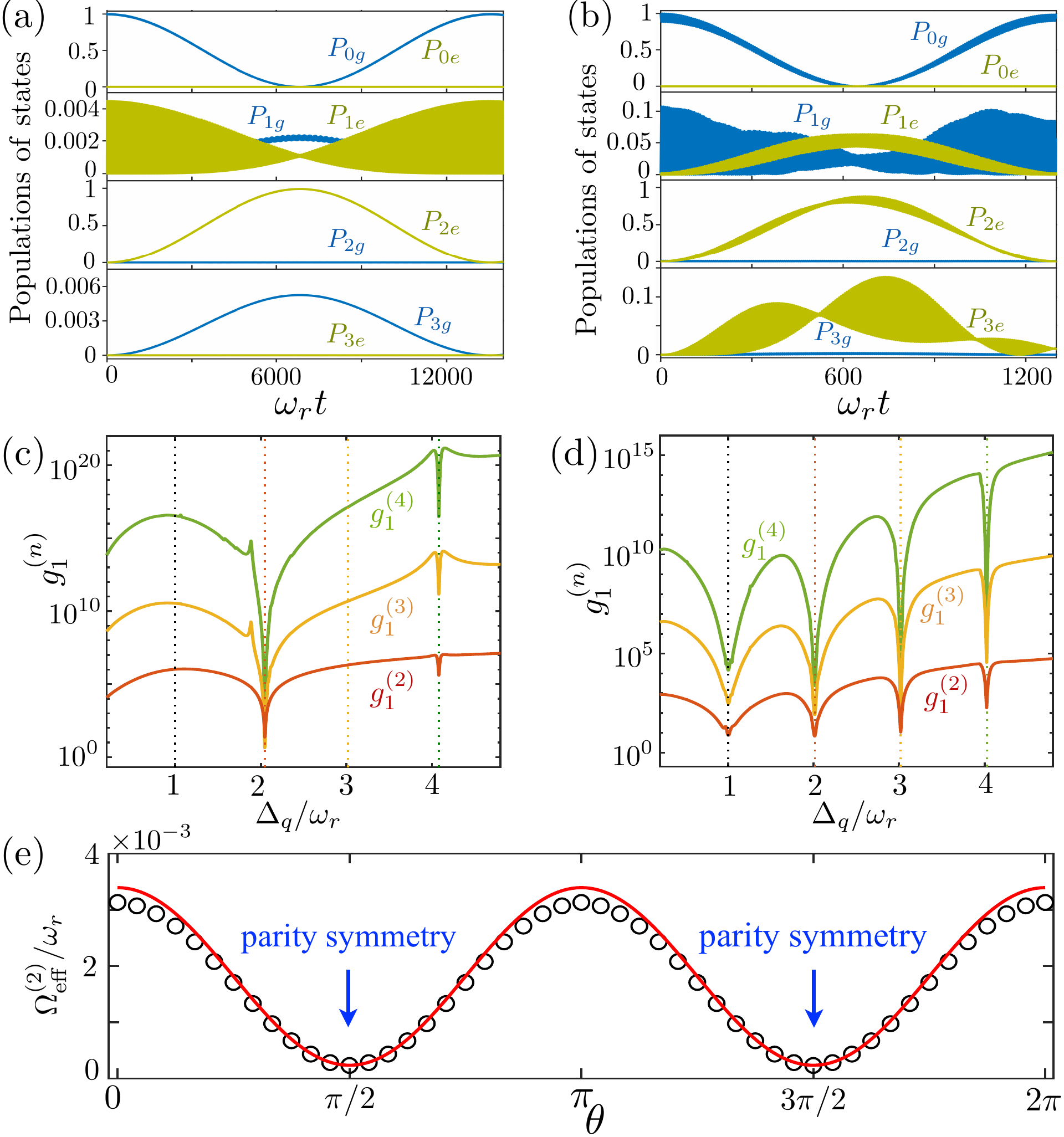}
  \caption{(a),(b) Time evolution of the populations of system states  $P_{nl}(t)=|\langle n,l|\psi(t)\rangle|^2$ ($l=e,g$ and $n=0,1,2,3$) at two-photon resonances in the absence of dissipation for (a) $\theta=\pi/2$ and (b) $\theta=\pi/6$.   (c),(d) Equal-time $n$th-order correlation functions $g^{(n)}_1$ as a function of $\Delta_q/\omega_r$ for (c) $\theta=\pi/2$ and (d) $\theta=\pi/6$ in the presence of dissipation. (e) The effective two-photon transition rate obtained by analytical solution $\Omega_{\rm eff}^{(2)}$ (red curve) and full-numerical simulation (black circles). Other system parameters are the same as in Fig.\,\ref{fig1}(b).
  }\label{fig2}
\end{figure}

To illustrate the above qualitative results, in Figs.\,\ref{fig2}(a) and \ref{fig2}(b), we present the super-Rabi oscillations $|0,g\rangle\leftrightarrow|2,e\rangle$ in the absence of dissipation. Comparing Figs.\,\ref{fig2}(a) and \ref{fig2}(b), it is shown that the values of $P_{1e}$ and $P_{3e}$ are suppressed by almost 2 orders in the case of $\theta=\pi/2$, and then they are ignorable.  This leads to that the almost perfect super-Rabi oscillation $|0,g\rangle\leftrightarrow|2,e\rangle$ is obtained in Fig.\,\ref{fig2}(a). Moreover, we also obtain the analytical rate of two-photon super-Rabi oscillation $\Omega_{\rm eff}^{(2)}=(\sqrt{2}\Omega\lambda^2/2)\{\sin^2\theta [1/(\omega_q+\omega_r)+1/(\omega_q-\omega_r)] [1/2\omega_r-1/(\omega_q+\omega_r)]+2\cos^2\theta/\omega_r^2\}$, by employing the unitary transformations $U_1$, $U_2$, $U_3$ and safely neglecting the direct transitions from $|0, g\rangle$~\cite{supp}. Figure\,\ref{fig2}(e) shows a very good agreement between $\Omega_{\rm eff}^{(2)}$ and the fully numerical simulation, which demonstrates the validity of our approximation.

\emph{Parity-symmetry-protected bundle emission}.---As a trigger of quantum emission, dissipation has to be considered here. In the ultrastrong coupling regime, the standard quantum master equation fails to correctly describe the dynamics of the system, since the qubit and cavity form an inseparable system\,\cite{F. Beaudoin}. One needs to account for the renormalization of the dissipator arising from the underlying system-bath formalism\,\cite{Malekakhlagh2019}, then the master equation is given by\,\cite{supp}
\begin{equation}\label{eq2}
\frac {d\rho}{dt}=i[\rho,H]+\kappa\mathcal{L}[X]+\gamma\mathcal{L}[D],
\end{equation}
where $\mathcal{L}[C]=(2C \rho C^\dag-\rho C^\dag C -C^\dag C \rho)/2$ is the renormalized dissipator, and $\kappa$ ($\gamma$) is the decay rate of cavity (qubit). The operators $X=\sum_{n,m>n}\langle  \psi_n|(a+a^\dag)| \psi_m\rangle|\psi_n\rangle \langle \psi_m|$ and $D=\sum_{n,m>n}\langle  \psi_n|(\sigma+\sigma^\dag)| \psi_m\rangle|\psi_n\rangle \langle \psi_m|$, in which $|\psi_n\rangle$ is the eigenstate of $\tilde{H}_{R}$.  The periodic drive is included in Eq.\,(\ref{eq2}) by time-dependent Hamiltonian $H$, and Eq.\,(\ref{eq2}) can be directly calculated by QuTip\,\cite{qutip}.
\begin{figure}
  \centering
  \includegraphics[width=8.7cm]{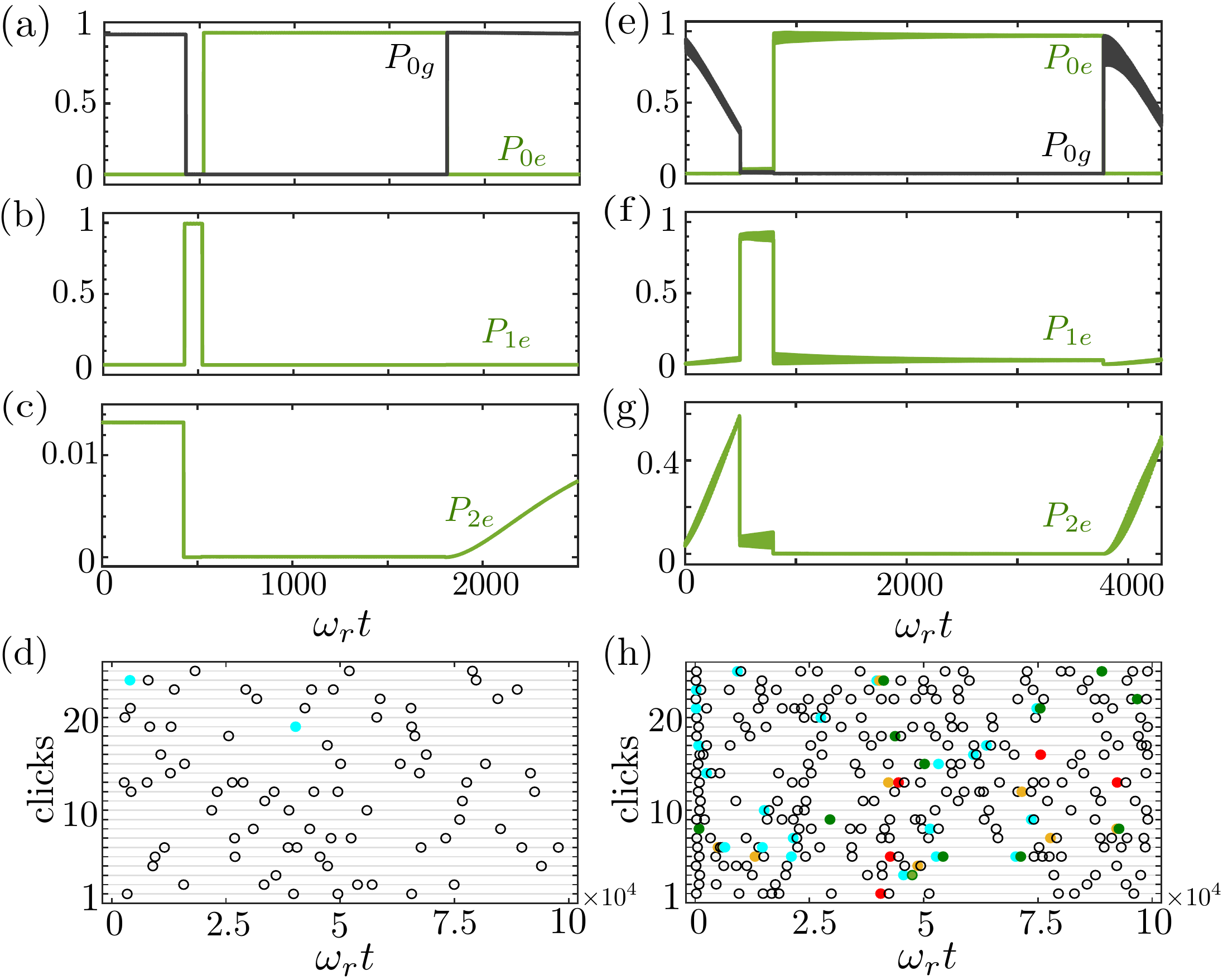}
  \caption{ A tiny fraction of a quantum trajectory (top) and the clicks over $25$ quantum trajectories (bottom) by quantum Monte Carlo  simulation in the two-photon emission regimes for (a)--(d) $\theta=\pi/2$ and (e)--(h) $\theta=\pi/6$. The lines in (a)--(c) and (e)--(g) represent the population dynamics of the different states $P_{nl}(t)$, and system  parameters are the same as in Fig.\,\ref{fig1}(b). Black, red and blue clicks represent the emission events $|2,e\rangle \to |1,e\rangle \to |0,e\rangle \to |0,g\rangle$,  $|2,e\rangle \to |2,g\rangle \to |1,g\rangle \to |0,g\rangle$ and $|2,e\rangle \to |1,e\rangle \to |1,g\rangle \to |0,g\rangle$, respectively. Green and yellow clicks represent the single-photon emission and other destructed emissions, respectively.}\label{fig3}
\end{figure}

Dissipation enables the multiphoton super-Rabi oscillation to be the bundles of strongly correlated photons outside of the system. According to the input-output theory, the strong correlation of emitted photons is now decided by the defined equal-time $n$th-order correlation $g^{(n)}_1\!\!=\!\!\langle X^{\dagger n} X^n\rangle/\langle X^{\dagger} X\rangle^n$\,\cite{supp, A. Ridolfo,Glauber1963}. As shown in Figs.\,\ref{fig2}(c) and \ref{fig2}(d), the dips inside the bunching peaks for all order of the correlations clearly associate to the multiphoton resonances. What appears in these resonances is the dips rather than the super-bunching peaks, which means that the system has entered into the emission regime of photon bundle. From Fig.\,\ref{fig2}(c), one can see that the dips associated with the multiphoton resonances can only occur at $\Delta_q\approx2n\omega_r$ in the case of $\theta=\pi/2$. This is another evidence of the presence of parity-symmetry-protected super-Rabi oscillation, which only associates with the periodic generation of even correlated photons in cavity. Consequently the system only emits $2n$ correlated photons without the destruction of the adjacent off-resonance emission with odd photons, which ultimately leads to the significant enhancement of emission purity.  When the parity symmetry is broken, the dips are found in the position of $\Delta_q \approx n\omega_r$, as shown in Fig.\,\ref{fig2}(d), which means both even and odd photons can be correlated, and the bundle emission of any number of strongly correlated photons can be achieved in this case. Thus our proposal offers an effective method to control the even-odd property of emitted correlated photons by manipulating the parity symmetry of the system, which is feasible in the circuit QED systems~\cite{Liu2005,T. Niemczyk,F. Deppe,A. Fedorov}. 

To confirm the multiphoton bundle emission, we take a quantum Monte Carlo simulation to follow the individual trajectory of the system and record photon clicks whenever the system undergoes a quantum jump~\cite{C. S. Munoz1}. Figures\,\ref{fig3}(a)--\ref{fig3}(c) and \ref{fig3}(e)--\ref{fig3}(g) show a small fraction of a quantum trajectory during two-photon emission, by calculating the populations of the system to be in the states $|n,e/g\rangle$. The bundle emission processes are similar for both the cases of qubit-cavity interaction with and without parity symmetry. Initially, the system is mainly in the superposition state of $|0,g\rangle$ and $|2,e\rangle$, and the probability of the two-photon state is much larger than the single photon state. The system is therefore more likely to collapse into the two-photon state when the dissipation triggers a quantum collapse of the system wave function. This leads to the emission of the first photon and the system collapses to the one-photon state $|1,e\rangle$ with almost unit probability, which is subsequently emitted during the cavity lifetime, completing the two-photon bundle emission. Now two strongly correlated photons are emitted outside of the system in a very short temporal window and the system is left in the state $|0,e\rangle$. After direct emission of a single photon from the qubit flip $|0,e\rangle\to|0,g\rangle$, the system goes back to the starting point for repeating the processes of two-photon emission. In the next cycle, the system undergoes the same cascade emission of two strongly correlated photons, associated with a single photon emission with different frequency.  The intrinsic temporal structure of the emitted two-photon bundle corresponds to the cascade emission of Fock states, shown in the Supplemental Material\,\cite{supp}.

\begin{figure}
  \centering
  \includegraphics[width=8.6cm]{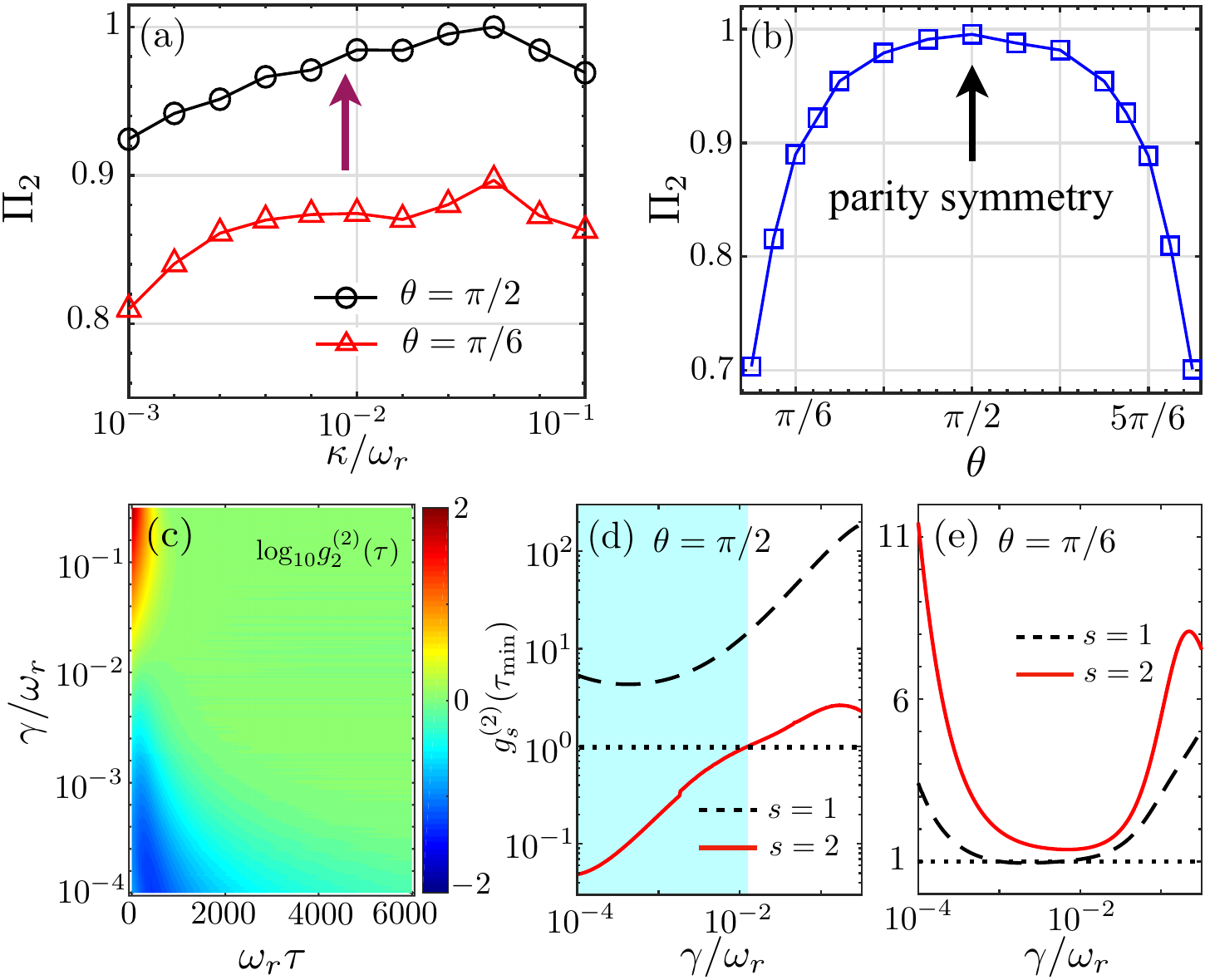}
  \caption{(a) Purity of two-photon emission $\Pi_2$ versus $\kappa/\omega_r$ for different values of $\theta$. (b) Purity $\Pi_2$ versus $\theta$ when $\kappa/\omega_r=10^{-1.6}$.  (c) Time-delay second-order bundle correlation $g^{(2)}_2(\tau)$ for $\theta=\pi/2$. (d),(e) Second-order correlation $g^{(2)}_s(\tau_{\rm min})$ with $s=1,2$ and $\tau_{\rm min}=1/\kappa$ versus $\gamma/\omega_r$ for (d) $\theta=\pi/2$ and (e) $\theta=\pi/6$. The shaded area in (d) corresponds to the regime of $g^{(2)}_1(\tau_{\rm min})>1$ and $g^{(2)}_2(\tau_{\rm min})<1$. The other system parameters are the same as in Fig.\,\ref{fig1}(b).  
  }\label{fig4}
\end{figure}

Comparing Figs.\,\ref{fig3}(b) and \ref{fig3}(f), it shows that the parity symmetry suppresses the single-photon populations before the emission process, which actually enhances the purity of two-photon bundle emission obtained by recording enough quantum trajectories~\cite{C. S. Munoz1}. To demonstrate this, in Figs.\,\ref{fig3}(d) and \ref{fig3}(h), we present photon clicks in the regime of two-photon emission, where each point indicates a single-photon or multiphoton emission event. The unexpected emission processes shown as the color clicks are significantly suppressed by the parity symmetry in the case of $\theta=\pi/2$, indicating the realization of parity-symmetry-protected $2n$-photon bundle emission. As a result, the purity $\Pi_2$ of bundle emission is significantly enhanced by the parity symmetry of system in a wide range of $\kappa/\omega_r$, as shown in Figs.\,\ref{fig4}(a) and \ref{fig4}(b). Here $\Pi_2=\mathcal{P}_2/\mathcal{P}_{\rm tot}$ is obtained by counting the number of two-photon bundle emission events $\mathcal{P}_2$ and the total number of all emission events $\mathcal{P}_{\rm tot}$ in the statistics of quantum trajectories \,\cite{supp}. Then, even in the regime of large cavity decay, almost perfect $2n$-photon bundle emission (here $n=1$) can be realized based on our proposal. Moreover, Figs.\,\ref{fig3}(d) and \ref{fig3}(h) also demonstrate that the rate of two-photon emission can be increased by breaking the parity symmetry of the system, due to the enhanced Rabi frequency in the case of $\theta\neq m\pi/2$ shown in Fig.\,\ref{fig2}(e).      

Although the function $g^{(n)}_1$ reveals the strong correlation of photons at multiphoton resonance, it fails to represent the actual multiphoton bundle correlation. To describe the statistical characteristics of bundles themselves, we numerically calculate the generalized second-order correlation of two-photon bundle\,\cite{C. S. Munoz1, Y. Chang, X.-L. Dong,supp}
\begin{align}\label{eq3}
g_2^{(2)}(\tau)=\frac{\langle X^{\dag 2}(0) X^{\dag 2}(\tau) X^{2}(\tau) X^{2}(0)\rangle}{\langle (X^{\dag 2} X^2)(0) \rangle\langle (X^{\dag 2} X^2)(\tau) \rangle}
\end{align}
where $\tau$ is the time delay and $\tau\geq\tau_{\rm min}=1/\kappa$. Here $\tau_{\rm min}$ can be approximately regarded as zero-time delay for the case of the bundle emission, since in the small-time windows of width $1/\kappa$ centered on zero, the bundle correlation function $g^{(2)}_2(\tau)$ is ill-defined (as such short times probe inside the bundle itself)\,\cite{C. S. Munoz1}. Figure\,\ref{fig4}(c) shows  how the bundle statistic $g^{(2)}_2(\tau)$ evolves from bunching for large $\gamma$ to antibunching for small $\gamma$, passing by the coherent emission with $g^{(2)}_2(\tau)=1$. Physically, the strong antibunching bundle is obtained in the case of the long-lived qubit (i.e., small $\gamma$), as it leaves enough time to separate successive emission events of the two-photon bundle. The threshold from bunching to antibunching can be enhanced by increasing $\lambda/\omega_r$ or $\Omega/\omega_r$\,\cite{supp}. The antibunching bundle corresponds to pairs of photons rather than larger bundles as $g^{(2)}_2(\tau_{\rm min})<1$, and the two photons among in a photon pair are bunched with $g^{(2)}_1(\tau_{\rm min})>1$, as shown in Fig.\,\ref{fig4}(d). It is also shown from Figs.\,\ref{fig4}(d) and \ref{fig4}(e) that the pair of photons can occur for a broad range of parameters when $\theta=\pi/2$, and never occur for $\theta= \pi/6$ due to the breaking of parity symmetry\,\cite{supp}.  

\emph{Discussion of experimental feasibility}.---Regarding the experimental implementations, a superconducting circuit is an ideal candidate for our proposal. We consider a circuit QED system where the $LC$ resonator, made of an inductor and a capacitor, is coupled to a flux qubit with energy gap $\Delta$ via the magnetic energy bias $\epsilon$~\cite{J. E. Mooij,  J. Q. You2007, Z. -L. Xiang, A. Fedorov, M. Stern, Gu2017}. The mixing angle $\theta$ can be controlled by adjusting the magnetic frustration of the qubit loop according to $\cos\theta=\epsilon/\sqrt{\Delta^2+\epsilon^2}$~\cite{A. Fedorov, M. Stern}.  
The qubit is driven by coupling to the current $I_d$ of the 1D transmission line through the dipole matrix element $\mu$, and the diving amplitude $\Omega=\mu I_d$\,\cite{O. Astafiev, M. J. Schwarz, X. Wang}. The correlation function of microwave photons can be measured by  employing  quadrature amplitude detectors\,\cite{Bozyigit2011, Lang2011}. The odd-even property of the emitted bundle of photons could be identified via measuring the excitation of an auxiliary flux qubit, when the emitted photons are scanned into this qubit\,\cite{F. Deppe}.  In this design, we theoretically predict that close to $100\%$ two-photon emission could be achieved with feasible experimental parameters ($\omega_r/2\pi=5{\rm ~GHz}$, $\omega_q/2\pi=25{\rm ~GHz}$,  $\lambda/2\pi=1{\rm ~GHz}$, $\Omega/2\pi=0.3{\rm ~GHz}$, $\gamma/2\pi=0.5{\rm ~MHz}$ and $\kappa/2\pi=200{\rm ~MHz}$)\,\cite{Kockum2019, FornDiaz2019,Gu2017}. 

\emph{Conclusion}.---We have investigated the multiphoton correlated emission in a cavity QED system under the ultrastrong coupling regime. The parity-symmetry-protected bundle emission is proposed in the driven Rabi model by combining the laser driven qubit flip and the state transitions in the odd parity chains induced by Rabi interaction. As a result, the parity symmetry only enables the realization of the $2n$-photon bundle emission. The corresponding purity is significantly enhanced over a wide range of parameters, comparing with the case of parity symmetry breaking. Our proposal promises almost perfect $2n$-photon bundle emission, and it extends the multiphoton bundle emission to the ultrastrong coupling regime. It is also fundamentally interesting in combining the symmetry theory.   

This work is supported by the National Key Research and Development Program of China Grant No. 2016YFA0301203 and the National Science Foundation of China (Grants No.\,11822502, No.\,11974125, and No.\,11875029).


\begin{thebibliography}{4}
\bibitem{H. Weyl} H.\,Weyl, \emph{Theory of Groups and Quantum Mechanics} (Methuen, London, 1931).

\bibitem{S. Singh} S.\,Singh, R.\,N.\,C. Pfeifer, and G.\,Vidal, Tensor network decompositions in the presence of a global symmetry, Phys. Rev. A \textbf{82}, 050301(R) (2010).

\bibitem{D. Braak2011} D. Braak, Integrability of the Rabi Model, Phys. Rev. Lett. \textbf{107}, 100401 (2011).

\bibitem{J. Casanova2010} J. Casanova, G. Romero, I. Lizuain, J. J. Garc\'{i}a-Ripoll, and E. Solano, Deep Strong Coupling Regime of the Jaynes-Cummings Model, Phys. Rev. Lett. \textbf{105}, 263603 (2010).

\bibitem{Jaynes1963} E. T. Jaynes and F. W. Cummings, Comparison of quantum and semiclassical radiation theories with application to the beam maser,  IEEE Xplore \textbf{51}, 89 (1963).

\bibitem{G. Gunter} G. G\"{u}nter, A. A. Anappara, J. Hees, A. Sell, G. Biasiol, L. Sorna, S. D. Liberota, C. Ciuti, A. Tredicucci, A. Leitenstorfer, and R. Huber, Sub-cycle switch-on of ultrastrong light-matter interaction, Nature (London) \textbf{458}, 178 (2009).

\bibitem{Y. Todorov} Y. Todorov, A. M. Andrews, R. Colombelli, S. De Liberato, C. Ciuti, P. Klang, G. Strasser, and C. Sirtori, Ultrastrong Light-Matter Coupling Regime with Polariton Dots, Phys. Rev. Lett. \textbf{105}, 196402 (2010).

\bibitem{P. Forn-Diaz} P. Forn-D\'{i}az, J. Lisenfeld, D. Marcos, J. J. Garc\'{i}a-Ripoll, E. Solano, C. J. P. M. Harmans, and J. E. Mooij, Observation of the Bloch-Siegert Shift in a Qubit-Oscillator System in the Ultrastrong Coupling Regime, Phys. Rev. Lett. \textbf{105}, 237001 (2010).

\bibitem{T. Schwartz} T. Schwartz, J. A. Hutchison, C. Genet, and T. W. Ebbesen, Reversible Switching of Ultrastrong Light-Molecule Coupling, Phys. Rev. Lett. \textbf{106}, 196405 (2011).

\bibitem{A. J. Hoffman} A. J. Hoffman, S. J. Srinivasan, S. Schmidt, L. Spietz, J. Aumentado, H. E. T\v{s}reci, and A. A. Houck, Dispersive Photon Blockade in a Superconducting Circuit, Phys. Rev. Lett. \textbf{107}, 053602 (2011).

\bibitem{C. M. Wilson} C. M. Wilson, G. Johansson, A. Pourkabirian, M. Simoen, J. R. Johansson, T. Duty, F. Nori, and P. Delsing, Observation of the dynamical Casimir effect in a superconducting circuit, Nature (London) \textbf{479}, 376 (2011).

\bibitem{G. Scalari} G. Scalari, C. Maissen, D. Tr\v{c}inkov\'{a}, D. Hagenm\"{u}ller, S. De Liberato, C. Ciuti, C. Reichl, D. Schuh, W. Wegscheider, M. Beck, and J. Faist, Ultrastrong Coupling of the cyclotron transition of a 2D electron gas to a THz metamaterial, Science \textbf{335}, 1323 (2012).

\bibitem{S. Gambino} S. Gambino, M. Mazzeo, A. Genco, O. D. Stefano, S. Savasta, S. Patan\`{e}, D. Ballarini, F. Mangione, G. Lerario, D. Sanvitto, and G. Gigli, Exploring light-matter interaction phenomena under ultrastrong coupling regime, ACS Photonics \textbf{1}, 1042 (2014).

\bibitem{P. Forn-Diaz2} P. Forn-D\'{\i}az, J. J. Garc\'{\i}a-Ripoll, B. Peropadre, J.-L. Orgiazzi, M. A. Yurtalan, R. Belyansky, C. M. Wilson, and A. Lupascu, Ultrastrong coupling of a single artificial atom to an electromagnetic continuum in the nonperturbative regime, Nat. Phys. \textbf{13}, 39 (2017).

\bibitem{FornDiaz2019} P.  Forn-D\'{i}az, L. Lamata, E. Rico, J. Kono, and E. Solano, Ultrastrong coupling regimes of light-matter interaction, Rev. Mod. Phys. \textbf{91}, 025005 (2019).

 \bibitem{Gu2017}  X. Gu, A. F. Kockum, A. Miranowicz, Y. -x. Liu, and F. Nori, Microwave photonics with superconducting quantum circuits, Phys. Rep. {\bf 718–719} 1–102 (2017).

 \bibitem{Kockum2019} A. F. Kockum, A. Miranowicz, S. De Liberato, S. Savasta, and F. Nori,  Ultrastrong coupling between light and matter, Nat. Rev. Phys. {\bf 1}, 19–40 (2019). 






\bibitem{C. Ciuti2} C. Ciuti, G. Bastard, and I. Carusotto, Quantum vacuum properties of the intersubband cavity polariton field, Phys. Rev. B \textbf{72}, 115303 (2005).

\bibitem{P. Nataf} P. Nataf and C. Ciuti, Vacuum Degeneracy of a Circuit QED System in the Ultrastrong Coupling Regime, Phys. Rev. Lett. \textbf{104}, 023601 (2010)

\bibitem{Romero2012} G. Romero, D. Ballester, Y. M. Wang, V. Scarani, and E. Solano, Ultrafast Quantum Gates in Circuit QED, Phys. Rev. Lett. \textbf{108}, 120501 (2012).



 \bibitem{Garziano2015} L. Garziano, R. Stassi, V. Macr\`{i}, A.F. Kockum, S. Savasta, and F. Nori,  Multiphoton quantum Rabi oscillations in ultrastrong cavity QED, Phys. Rev. A {\bf 92}, 063830 (2015).


\bibitem{Ma2020} Ken K. W. Ma, Multiphoton resonance and chiral transport in the generalized Rabi, Phys. Rev. A {\bf 102}, 053709 (2020).

\bibitem{R. Stassi} R. Stassi, A. Ridolfo, O. Di Stefano, M. J. Hartmann, and S. Savasta, Spontaneous Conversion from Virtual to Real Photons in the Ultrastrong-Coupling Regime, Phys. Rev. Lett. \textbf{110}, 243601 (2013).

 \bibitem{Kockum2017} A. F. Kockum, A. Miranowicz, V. Macr\`{i}, S. Savasta, and F. Nori,  Deterministic quantum nonlinear optics with single atoms and virtual photons, Phys. Rev. A {\bf 95}, 063849 (2017).

\bibitem{Malekakhlagh2019} M. Malekakhlagh and A. W. Rodriguez, Quantum Rabi Model with Two-Photon Relaxation, Phys. Rev. Lett. {\bf 122}, 043601 (2019)

\bibitem{V. Giovannetti} V. Giovannetti, S. Lloyd, and L. Maccone, Quantum Metrology, Phys. Rev. Lett. \textbf{96}, 010401 (2006).

\bibitem{M. DAngelo} M. D'Angelo, M.\,V. Chekhova, and Y. Shih, Two-Photon Diffraction and Quantum Lithography, Phys. Rev. Lett. \textbf{87}, 013602 (2001).

\bibitem{W. Denk} W. Denk, J. Strickler, and W. Webb, Two-photon laser scanning fluorescence microscopy, Science \textbf{248}, 73 (1990).

\bibitem{N. Horton} N. Horton, D. Wang, C. Kobat, F. Clark, C. Wise, and C.\,X.\,C. Schaffer,  {\it In vivo} three-photon microscopy of subcortical structures within an intact mouse brain, Nat. Photonics \textbf{7}, 205 (2013).

\bibitem{M. F. Maghrebi} M. F. Maghrebi, M. J. Gullans, P. Bienias, S. Choi, I. Martin, O. Firstenberg, M. D. Lukin, H. P. Bschler, and A. V. Gorshkov, Coulomb Bound States of Strongly Interacting Photons, Phys. Rev. Lett. \textbf{115}, 123601 (2015).

\bibitem{K. Jachymski} K. Jachymski, P. Bienias, and H. P. B\"{u}chler, Three-Body Interaction of Rydberg Slow-Light Polaritons, Phys. Rev. Lett. {\bf 117}, 053601 (2016).

\bibitem{Q.-Y. Liang} Q.-Y. Liang, A. V. Venkatramani, S. H. Cantu1, T. L. Nicholson, M. J. Gullans, A. V. Gorshkov, J. D. Thompson, C. Chin, M. D. Lukin, and V. Vuleti\'{c}, Observation of three-photon bound states in a quantum nonlinear medium, Science \textbf{359}, 783 (2018).

\bibitem{Deng2020} Y.  G. Deng, T. Shi, S. Yi, Motional \emph{n}-phonon bundle states of a trapped atom with clock transitions, Photon. Res. {\bf 9}, 1289 (2021).

\bibitem{J. S. Douglas} J. S. Douglas, T. Caneva, and D. E. Chang, Photon Molecules in Atomic Gases Trapped Near Photonic Crystal Waveguides, Phys. Rev. X \textbf{6}, 031017 (2016).

\bibitem{A. Gonzalez-Tudela2} A. Gonz\'{a}lez-Tudela, V. Paulisch, H. J. Kimble, and J. I. Cirac, Efficient Multiphoton Generation in Waveguide Quantum Electrodynamics, Phys. Rev. Lett. \textbf{118}, 213601 (2017).

\bibitem{Y. Ota} Y. Ota, S. Iwamoto, N. Kumagai, and Y. Arakawa, Spontaneous Two-Photon Emission from a Single Quantum Dot, Phys. Rev. Lett. \textbf{107}, 233602 (2011).

\bibitem{C. S. Munoz1} C. S. Mu\~{n}oz, E. del Valle, A. G. Tudela, K. M\"{u}ller, S. Lichtmannecker, M. Kaniber, C. Tejedor, J. Finley, and F. Laussy, Emitters of N-photon bundles, Nat. Photonics \textbf{8}, 550 (2014).

\bibitem{Y. Chang} Y. Chang, A. Gonz\'{a}lez-Tudela, C. S. Mu\~{n}oz, C. Navarrete-Benlloch, and T. Shi, Deterministic Down-Converter and Continuous Photon-Pair Source within the Bad-Cavity Limit, Phys. Rev. Lett. \textbf{117}, 203602 (2016).

\bibitem{C. S. Munoz2} C. S. Mu\~{n}oz, F. P. Laussy, E. del Valle, C. Tejedor, and A. Gonz\'{a}lez-Tudela, Filtering multiphoton emission from state-ofthe-art cavity quantum electrodynamics, Optica \textbf{5}, 14 (2018).

\bibitem{W. Qin} W. Qin, V. Macrì, A. Miranowicz, S. Savasta, and F. Nori, Emission of photon pairs by mechanical stimulation of the squeezed vacuum, Phys. Rev. A \textbf{100}, 062501 (2019).

\bibitem{K. J. Satzinger} K.\,J. Satzinger, Y.\,P. Zhong, H.\,-S. Chang, G.\,A. Peairs, A. Bienfait, M\,-H. Chou, A.\,Y. Cleland, C.\,R. Conner, \~{E}. Dumur, J. Grebel, I. Gutierrez, B.\,H. November, R.\,G. Povey, S.\,J. Whiteley, D.\,D. Awschalom, D.\,I. Schuster, and A.\,N. Cleland, Quantum control of surface acoustic-wave phonons, Nature (London) \textbf{563}, 661 (2018).

\bibitem{D. V. Strekalov} D. V. Strekalov, A bundle of photons, please,  Nat. Photonics \textbf{8}, 500 (2014).


\bibitem{Y. W. Chu} Y.\,W. Chu, P. Kharel, T. Yoon, L. Frunzio, P.\,T. Rakich, and  R.\,J. Schoelkopf, Creation and control of multi-phonon Fock states in a bulk acoustic-wave resonator, Nature (London) \textbf{563}, 666 (2018).

\bibitem{D. J. Gauthier} D.\,J. Gauthier, Q.\,L. Wu, S.\,E. Morin, and T.\,W. Mossberg, Realization of a Continuous-Wave, Two-Photon Optical Laser, Phys. Rev. Lett. \textbf{68}, 464 (1992).


\bibitem{Q. Bin2} Q. Bin, X. -Y. L\"{u}, T. -S. Yin, Y. Li, and Y. Wu, Collective radiance effects in the ultrastrong-coupling regime, Phys. Rev. A \textbf{99}, 033809 (2019).






\bibitem{Liu2005} Y. -x. Liu, J. Q. You, L. F. Wei, C. P. Sun, and F. Nori, Optical Selection Rules and Phase-Dependent Adiabatic State Control in a Superconducting Quantum Circuit, Phys. Rev. Lett. {\bf 95}, 087001 (2005).

\bibitem{T. Niemczyk} T. Niemczyk, F. Deppe, H. Huebl, E. P. Menzel, F. Hocke, M. J. Schwarz, J. J. Garc\'{i}a-Ripoll, D. Zueco, T. H\"{u}mmer,	E. Solano, A. Marx, and R. Gross, Circuit quantum electrodynamics in the ultrastrong-coupling regime, Nat. Phys. \textbf{6}, 772 (2010).

\bibitem{F. Deppe} F. Deppe, M. Mariantoni, E. P. Menzel, A. Marx, S. Saito, K. Kakuyanagi, H. Tanaka, T. Meno, K. Semba, H. Takayanagi, E. Solano, and R. Gross, Two-photon probe of the Jaynes-Cummings model and controlled symmetry breaking in circuit QED, Nat. Phys. \textbf{4}, 686 (2008).


\bibitem{A. Fedorov} A. Fedorov, A. K. Feofanov, P. Macha, P. Forn-D\'{\i}az, C. J. P. M. Harmans, and J. E. Mooij, Strong Coupling of a Quantum Oscillator to a Flux Qubit at Its Symmetry Point, Phys. Rev. Lett. \textbf{105}, 060503 (2010).

\bibitem{supp} See Supplemental Material at URL for the analytical calculation of the super-Rabi oscillation frequency, derivation of  the master equation in dressed picture, discussion of the intrinsic temporal structure of emitted multiphoton bundle, discussion of quantum correlations of multiphoton bundles, and purity of the multiphoton bundle emission, which includes Refs.\,\cite{Zueco2009RK, Chen2017WL, Gardiner2000Zoller, Ma2015Law, L. Garziano,Walls2008Milburn}.

\bibitem{Zueco2009RK} D. Zueco, G. M. Reuther, S. Kohler, and P. H\"{a}nggi, Qubit-oscillator dynamics in the dispersive regime: Analytical theory beyond the rotating-wave approximation, Phys. Rev. A {\bf 80}, 033846 (2009).

\bibitem{Chen2017WL} Z. Chen, Y. M. Wang, T. F. Li, L. Tian, Y. Y. Qiu, K. Inomata, F. Yoshihara, S. Y. Han, F. Nori, J. S. Tsai, and J. Q. You, Single-photon-driven high-order sideband transitions in an ultrastrongly coupled circuit-quantum-electrodynamics system, Phys. Rev. A. \textbf{96}, 012325 (2017).

\bibitem{Gardiner2000Zoller} C. W. Gardiner and P. Zoller, \emph{Quantum Noise}, 2nd ed. (Springer, Berlin, 2000).

\bibitem{Ma2015Law} K. K. W. Ma and C. K. Law, Three-photon resonance and adiabatic passage in the large-detuning Rabi model, Phys. Rev. A. {\bf 92}, 023842 (2015).

\bibitem{L. Garziano} L. Garziano, V. Macr\`{\i}, R. Stassi, O. Di Stefano, F. Nori, and S. Savasta, One Photon Can Simultaneously Excite Two or More Atoms, Phys. Rev. Lett. \textbf{117}, 043601 (2016).

\bibitem{Walls2008Milburn}  D. F. Walls and G. J. Milburn, \emph{Quantum Optics} (Springer, New York, 2008).




\bibitem{F. Beaudoin} F. Beaudoin, J. M. Gambetta, and A. Blais, Dissipation and ultrastrong coupling in circuit QED, Phys. Rev. A \textbf{84}, 043832 (2011).

\bibitem{qutip} J. R. Johansson, P. D. Nation, and F. Nori, QuTiP: An open- source Python framework for the dynamics of open quantum systems, Comput. Phys. Commun. {\bf 183}, 1760 (2012).

\bibitem{A. Ridolfo} A. Ridolfo, M. Leib, S. Savasta, and M. J. Hartmann, Photon Blockade in the Ultrastrong Coupling Regime, Phys. Rev. Lett. \textbf{109}, 193602 (2012).

\bibitem{Glauber1963}  R. J. Glauber, The quantum theory of optical coherence, Phys. Rev. \textbf{130}, 2529 (1963).

\bibitem{X.-L. Dong} X. -L. Dong and P. -B. Li, Multiphonon interactions between nitrogen-vacancy centers and nanomechanical resonators, Phys. Rev. A \textbf{100}, 043825 (2019).


\bibitem{J. E. Mooij} J. E. Mooij, T. P. Orlando, L. Levitov, L. Tian, C. H. Van der Wal, and S. Lloyd, Josephson persistent-current qubit, Science {\bf 285}, 1036 (1999).

\bibitem{J. Q. You2007} J. Q. You, Y. X. Liu, C. P. Sun, and F. Nori, Persistent single-photon production by tunable on-chip micromaser with a superconducting quantum circuit, Phys. Rev. B {\bf 75}, 104516 (2007).

\bibitem{Z. -L. Xiang} Z. -L. Xiang, S. Ashhab, J. Q. You, and F. Nori, Hybrid quantum circuits: Superconducting circuits interacting with other quantum systems, Rev. Mod.  Phys. {\bf 85}, 623 (2013).

\bibitem{M. Stern} M. Stern, G. Catelani, Y. Kubo, C. Grezes, A. Bienfait, D. Vion, D. Esteve, and P. Bertet, Flux Qubits with Long Coherence Times for Hybrid Quantum Circuits, Phys. Rev. Lett. {\bf 113}, 123601 (2014).

\bibitem{M. J. Schwarz} M. J. Schwarz, Gradiometric tunable-gap flux qubits in a circuit QED architecture, Ph.D. thesis, Technische Universit\"{a}t M\"{u}nchen, Mt\"{u}nchen, 2015.

\bibitem{X. Wang} X. Wang, A. Miranowicz, H. -R. Li, and F. Nori, Hybrid quantum device with a carbon nanotube and a flux qubit for dissipative quantum engineering, Phys. Rev. B {\bf 95}, 205415 (2017).

\bibitem{O. Astafiev} O. Astafiev, A. M. Zagoskin, A. A. Abdumalikov, Yu. A. Pashkin, T. Yamamoto, K. Inomata, Y. Nakamura, and J. S. Tsai, Resonance fluorescence of a single artificial atom, Science {\bf 327}, 840 (2010).

\bibitem{Bozyigit2011}  D. Bozyigit, C. Lang, L. Steffen, J. M. Fink, C. Eichler, M. Baur, R. Bianchetti, P. J. Leek, S. Filipp, M. P. da Silva, A. Blais, and A. Wallraff, Antibunching of microwave-frequency photons observed in correlation measurements using linear detectors, Nat. Phys. {\bf 7}, 154 (2011). 

\bibitem{Lang2011}  C. Lang, D. Bozyigit, C. Eichler, L. Steffen, J. M. Fink, A. A. Abdumalikov, Jr., M. Baur, S. Filipp, M. P. da Silva, A. Blais, and A. Wallraff, Observation of Resonant Photon Blockade at Microwave Frequencies Using Correlation Function Measurements, Phys. Rev. Lett. {\bf 106}, 243601  (2011). 




\end{thebibliography}
\end{document}